\documentclass[a4paper,11pt]{article}
\pdfoutput=1
\usepackage{jheppub} 
\usepackage[T1]{fontenc} 
\usepackage[english]{babel}
\usepackage[latin1]{inputenc}
\usepackage{amsmath,mathtools}
\usepackage{amsfonts}
\usepackage{amssymb}
\usepackage{wrapfig}
\usepackage{color}
\usepackage{float}
\usepackage{url}
\usepackage{mciteplus}
\usepackage{dsfont}
\usepackage{ifpdf}
\usepackage{epsfig}
\usepackage{xcolor}
\usepackage{hyperref}
\usepackage{cancel}
\usepackage{arydshln}

\allowdisplaybreaks

\DeclareMathOperator{\Jt}{\widetilde{\mathnormal{J}}}

\DeclareMathOperator{\dt}{\widetilde{\mathnormal{d}}}

\DeclareMathOperator{\Rgod}{\mathnormal{R}_\mathnormal{g}\o\mathnormal{d}}
\DeclareMathOperator{\Rgodt}{\mathnormal{R}_\mathnormal{g}\o\widetilde{\mathnormal{d}}}
\renewcommand{\a}{\alpha}
\renewcommand{\b}{\beta}
\newcommand{\g}{\gamma}
\newcommand{\G}{\Gamma}
\renewcommand{\k}{\kappa}
\renewcommand{\l}{\lambda}

\newcommand{\M}{\mathcal{M}}
\newcommand{\N}{\mathcal{N}}
\newcommand{\K}{\mathcal{K}}
\newcommand{\norm}[1]{\left\|#1\right\|}
\renewcommand{\o}{\circ}
\newcommand{\Adg}{\text{Ad}_g}
\newcommand{\Adgi}{\text{Ad}_g^{-1}}
\newcommand{\Pm}{\mathnormal{P}_{-}}
\newcommand{\Str}{\text{Str}}
\newcommand{\tr}{\text{tr}}

\newcommand{\Rg}[1]{\mathnormal{R}_\mathnormal{g}\left(#1\right)}

\newcommand{\dds}{\mathnormal{d}^2\sigma\,}
\newcommand{\eRg}{\eta R_g}
\renewcommand{\d}{\delta}

\newcommand{\diag}{\text{diag}}

\newcommand{\algg}{\mathfrak{g}}
\newcommand{\alggb}{\mathfrak{g}_{\text{b}}}

\newcommand{\so}{\mathfrak{so}}
\newcommand{\su}{\mathfrak{su}}
\renewcommand{\u}{\mathfrak{u}}

\newcommand{\psu}{\mathfrak{psu}}
\newcommand{\algi}[1]{\mathfrak{g}^{(#1)}}
\newcommand{\Ai}[1]{A^{(#1)}}

\newcommand{\Z}[1]{\mathbb{Z}_{#1}}

\newcommand{\Em}[1]{E^{#1}}

\newcommand{\Cmn}[2]{C_{#1}^{\;\;#2}}

\newcommand{\Lmn}[2]{\Lambda_{#1}^{\;\;#2}}
\newcommand{\dmn}[2]{\delta_{#1}^{\;\;#2}}
\newcommand{\Pb}[1]{P_2\left(#1\right)}
\newcommand{\jm}[1]{j^{#1}}

\newcommand{\Oi}{\mathcal{O}^{-1}}

\newcommand{\x}[1]{x^{#1}}

\newcommand{\xm}{x_{-}}
\newcommand{\xp}{x_{+}}

\newcommand{\dx}[1]{dx^{#1}}
\newcommand{\dxs}[1]{dx_{#1}^2}

\newcommand{\dxm}{dx_{-}}
\newcommand{\dxp}{dx_{+}}

\newcommand{\dxps}{dx_{+}^2}

\newcommand{\AdS}[1]{AdS_{#1}}
\newcommand{\Sph}[1]{S^{#1}}
\newcommand{\CP}[1]{\mathbb{CP}^{#1}}
\newcommand{\T}[1]{T^{#1}}

\newcommand{\Sch}[1]{Sch_{#1}}

\newcommand{\EK}[1]{\textbf{K}_{#1}}

\newcommand{\EH}{\textbf{H}}
\newcommand{\EM}{\textbf{M}}
\newcommand{\ET}[1]{\textbf{T}_{#1}}
\newcommand{\EX}[1]{\textbf{X}_{#1}}
\newcommand{\EY}[1]{\textbf{Y}_{#1}}

\newcommand{\SOM}[1]{\textbf{M}_{#1}}
\newcommand{\Ep}[1]{\textbf{p}_{#1}}
\newcommand{\ED}{\textbf{D}}
\newcommand{\Epm}{\textbf{p}_{-}}
\newcommand{\Epp}{\textbf{p}_{+}}

\newcommand{\es}[1]{\sigma_{#1}}

\renewcommand{\th}[1]{\theta_{#1}}
\newcommand{\dth}[1]{d\theta_{#1}}

\newcommand{\ph}[1]{\phi_{#1}}

\newcommand{\dph}[1]{d\phi_{#1}}

\newcommand{\Et}[1]{\eta_{#1}}

\newcommand{\eabc}[3]{\epsilon_{#1#2}^{\quad#3}}

\newcommand{\CC}{\mathbb{C}}

\newcommand{\gab}[1]{\gamma^{#1}}
\newcommand{\eab}[1]{\varepsilon^{#1}}

\newcommand{\Lap}{\nabla^2}

\title{\boldmath 
Yang-Baxter Deformations of the $\AdS{5}\times\T{1,1}$ Superstring and their Backgrounds}

\author{Laura Rado,}
\author{Victor O. Rivelles}
\author{and Renato Sánchez}

\affiliation{Instituto de F\'{\i}sica, Universidade de S\~{a}o Paulo \\ Rua do Mat\~{a}o Travessa 1371, 05508-090 S\~{a}o Paulo, SP. Brazil}

\emailAdd{laura@if.usp.br}
\emailAdd{rivelles@fma.if.usp.br}
\emailAdd{renato@if.usp.br}

\abstract{
We consider three-parameter Yang-Baxter deformations of the $\AdS{5}\times\T{1,1}$ superstring for abelian $r$-matrices which are solutions of the classical Yang-Baxter equation. We find the NSNS fields of two new backgrounds which are dual to the dipole deformed Klebanov-Witten gauge theory and to the nonrelativistic Klebanov-Witten gauge theory with Schr\"odinger symmetry. 
} 

\makeatletter
\gdef\@fpheader{}
\makeatother

\begin{document}
\maketitle
\flushbottom
\section{Introduction}
The AdS/CFT correspondence conjectures that certain gauge theories have a dual description in terms of string theories. The first case of the AdS/CFT correspondence states that $\N=4$ supersymmetric Yang-Mills theory on a four-dimensional flat spacetime is dual to type IIB superstring theory propagating in $\AdS{5}\times\Sph{5}$ \cite{Maldacena1999a}. 
One of the most important features of the AdS/CFT correspondence is its integrability which   
in the string theory side is associated to the existence of a Lax connection ensuring the existence of an infinite number of conserved charges.
In the case of $\AdS{5}\times\Sph{5}$ superstring, the theory is described by a $\sigma$-model on the supercoset $\frac{PSU(2,2|4)}{SO(1,4)\times SO(5)}$ \cite{Metsaev1998} and the $\Z{4}$-grading of the $\psu(2,2|4)$ superalgebra is an essential ingredient to get a Lax connection 
\cite{Bena2004}. The same happens for the  $\AdS{4}\times\CP{3}$ superstrings \cite{Aharony2008}, partially described by the supercoset $\frac{UOSp(2,2|6)}{SO(1,3)\times U(3)}$ \cite{Arutyunov2008,Stefanski2009}, which also has $\Z{4}$-grading and is integrable \cite{Arutyunov2008}.

Another way to get integrable theories is to start with an integrable model and then deformed it in such a way that integrability is preserved. This is accomplished by introducing $r$-matrices that satisfy the Yang-Baxter equation \cite{Klimcik2002}. 
When applied to the $\AdS{5}\times\Sph{5}$ case \cite{Delduc2014,Delduc2014a} the superstring will propagate on what is called a 
$\eta$-deformed background 
which is not a solution of the standard type IIB supergravity equations \cite{Arutyunov2014,Arutyunov2015}, 
leading to the proposal of generalized supergravities \cite{Arutyunov2016,Wulff2016}. 

Deformations based on r-matrices that satisfy the classical Yang-Baxter equation (CYBE) can also be considered \cite{Matsumoto2015}. When applied to superstrings in $\AdS{5}\times\Sph{5}$ \cite{Kawaguchi2014,Matsumoto2014,Matsumoto2014a,Matsumoto2015a,Kyono2016} they generate type IIB supergravity backgrounds like the Lunin-Maldacena-Frolov \cite{Lunin2005,Frolov2005}, Hashimoto-Itzhaki-Maldacena-Russo \cite{Hashimoto1999,Maldacena1999} \footnote{This background was also obtained as the $\eta\rightarrow0$ limit of the $\eta$-deformed $\AdS{5}\times\Sph{5}$ background \cite{Arutyunov2014} after a rescaling \cite{Arutyunov2015,Hoare2016b}.} and Schr\"odinger spacetimes \cite{Maldacena2008,Herzog2008,Adams2008,Bobev2009} which were previously obtained by TsT transformations \cite{Osten2017}
\footnote{It should be remarked that these deformations are generated by abelian r-matrices which, from the TsT side, involve commuting isometries. More general r-matrices, however, are associated to non-abelian T-dualities \cite{Matsumoto2015c,Tongeren2015,Hoare2016,Borsato2016,Borsato2017}.}.
In terms of TsT transformations, these backgrounds can be obtained by considering two-tori with directions either along the brane or transverse to it, or with a direction along the brane and the other transverse to it \cite{Imeroni2008}. In the first case, the two-torus along the brane will be generated by momenta operators, which introduce noncommutativity in the dual field theory \footnote{It is argued in \cite{Tongeren2016,Tongeren2017,Araujo2017a,Araujo2018} that noncommutativity, in general, can be introduced by considering conformal twists, or in terms of Yang-Baxter deformations by taking generators of the conformal algebra.}, while the two-torus in the transverse space to the brane will be along the $U(1)$ directions generated by the Cartan generators of the isometry group. These type of $U(1)\times U(1)$-deformations are called $\b$-deformations. Now, if the two-torus has directions one along the brane and the other transverse to it, taking a momentum and a Cartan generator, in the field theory side it leads to dipole field theories \cite{Dasgupta2000,Bergman2002}. Another possibility is to take a null direction along the brane and a $U(1)$ direction transverse to it. This generates gravity duals  of nonrelativistic field theories which have Schr\"odinger symmetry \cite{Hagen1972,Niederer1972,Son2008,Balasubramanian2008,Herzog2008,Guica2017}. The TsT procedure in this latter case is actually a TsssT transformation called Melvin twist \cite{Bobev2009,Bobev2009a}. In general, having more than one $U(1)$ direction in the transverse space allow us to construct the transverse two-torus in several ways so we  get different deformations or a combinations of them.
These Yang-Baxter deformations can also be applied to $\AdS{4}\times\CP{3}$ superstrings \cite{Negron2018,Rado2020} 
giving rise to gravity duals for the noncommutative, dipole and $\b$-deformed ABJM theory \cite{Imeroni2008}. It was also found an Yang-Baxter deformation that generates Schr\"odinger spacetimes which correspond to a family of gravity duals of nonrelativistic ABJM theory \cite{Rado2020}.

In this paper we will consider the duality between 
type IIB superstring theory in $\AdS{5}\times\T{1,1}$ and $\N=1$ $SU(N)\times SU(N)$ Yang-Mills theory in four dimensions, also known as Klebanov-Witten theory \cite{Klebanov1998}. The internal $\T{1,1}$ manifold has $SU(2)\times SU(2)\times U(1)_R$ symmetry, instead of the $SU(4)_R$ symmetry of the $\AdS{5}\times\Sph{5}$ case, leading to a less supersymmetric dual field theory. 
It has been argued that $\AdS{5}\times\T{1,1}$ is non-integrable since some wrapping string configurations present chaotic behavior \cite{Basu2011,Basu2011b}
\footnote{Notice, however, that a coset constructions for $\T{1,1}$ based on affine Gaudin models was found to be integrable \cite{Arutyunov2020}.
}. 
Even so, an integrable Yang-Baxter deformation of $\AdS{5}\times\T{1,1}$ was found \cite{Crichigno2014} which agrees with the gravity dual of the $\b$-deformed gauge theory obtained by TsT transformations \cite{Catal2006}.
In this paper we will discuss two other types of Yang-Baxter deformations generated by commuting $r$-matrices. One gives rise to a background dual to a three-parameter dipole deformed Klebanov-Witten gauge theory and the other is dual to a nonrelativistic Klebanov-Witten gauge theory on a Schr\"odinger spacetime.

This paper is organized as follows. 
In \autoref{cosetconstruction} we build the coset for $\AdS{5}\times\T{1,1}$ paying attention to the relevant subalgebras that will be used. 
In \autoref{news} we discuss the new backgrounds obtained by deforming $\AdS{5}\times\T{1,1}$. Finally, in \autoref{conclusions}, we discuss our results and present future perspectives.

\section{Coset construction of \texorpdfstring{$\AdS{5}\times\T{1,1}$}{AdS5xT11} \label{cosetconstruction}}
The Klebanov-Witten  gauge theory is obtained by putting $N$ $D3$-branes on the singularity of $M_{1,4}\times Y_6$, where $M_{1,4}$ is the four-dimensional Minkowski space and $Y_6$ a Ricci flat Calabi-Yau cone $C(X_5)$ with base $X_5$ \cite{Klebanov1998}. Near the horizon the geometry becomes $\AdS{5}\times X_5$, where $X_5$ is a compact Sasaki-Einstein manifold, i.e., an odd-dimensional Riemannian manifold such that its cone $C(X_5)$ is a Calabi-Yau flat manifold \cite{Sparks2011}. 
Taking $X_5$ as $\T{1,1}$ \footnote{This space belongs to a general class of Einstein spaces called $\T{p,q}$ \cite{Romans1985} described by the coset $SU(2)\times SU(2)/U(1)$, where the $U(1)$ is generated by $p\es{3}^L+q\es{3}^R$, where $\es{i}^L$ and $\es{j}^R$ are the generators of the left and right $SU(2)$'s, respectively.} only 1/4 of the supersymmetries are preserved so that we have $\N=1$ supersymmetry in four dimensions. 
The superpotential has a $SU(2)\times SU(2)\times U(1)$ symmetry, with $U(1)$ being part of the R-symmetry that gives the $\N=1$ supersymmetry, and $SU(2)\times SU(2)$ being a flavor symmetry which is not included in the $\N=1$ superconformal group in four dimensions $PSU(2,2|1)$ \cite{Ceresole2000,Ceresole1999,Ceresole2000a}. Thus, the full isometry group is $PSU(2,2|1)\times SU(2)\times SU(2)$.
The bosonic part of the superalgebra $\algg=\psu(2,2|1)$ on which we construct the $\sigma$-model is $\su(2,4)\otimes\u(1)$. The generators of $\psu(2,2|1)$
can be written as supermatrices which are formed by blocks that correspond to bosonic (diagonal) and fermionic (anti-diagonal) generators,
\begin{equation}
M_{(4|1)\times(4|1)}=\left(\begin{array}{c|c}
\so(2,4) & \overline{Q}\\ 
\hline
Q & \u(1)_R
\end{array}\right).
\end{equation}

The isometry group of $\AdS{5}\times\T{1,1}$ is given by the coset
\begin{equation}
\label{isometry}
\AdS{5}\times\T{1,1}\equiv\frac{SO(2,4)}{SO(1,4)}\times\frac{SU(2)\times SU(2)}{U(1)},
\end{equation}
which is not the bosonic part of any supercoset \cite{Kac1977,Nahm1978}. 
Besides that, the coset for $\T{1,1}$ does not lead to the standard Sasaki-Einstein metric for $\T{1,1}$. 
This happens because neither the bosonic subalgebra $\su(2)\otimes\u(1)$ nor the isometry group \eqref{isometry} captures the full isometries of the theory. All this can be overcome by extending the coset \eqref{isometry}  to \cite{Crichigno2014}
\begin{equation}
\label{isometry2}
\AdS{5}\times\T{1,1}\equiv\frac{SO(2,4)}{SO(1,4)}\times\frac{SU(2)\times SU(2)\times U(1)_R}{U(1)\times U(1)},
\end{equation}
where the $U(1)_R$ now appears as part of the global symmetries and a second $U(1)$ was added in order to preserve the number of parameters that describe the space.
Thus, in terms of this extended $\Z{2}$-graded algebra, the symmetric coset for $\AdS{5}\times \T{1,1}$ is taken as
\begin{equation}
\label{cosetT11}
\so(2,4)\oplus\su(2)\oplus\su(2)\oplus\u(1)=\overbrace{\left(\so(1,4)\oplus\u(1)\oplus\u(1)\right)}^{\algi{0}}\oplus\overbrace{\left(\frac{\so(2,4)\oplus\su(2)\oplus\su(2)\oplus\u(1)_R}{\so(1,4)\oplus\u(1)\oplus\u(1)}\right)}^{\algi{2}=\algg/\algi{0}}.
\end{equation}
The supermatrix has the block structure 
\begin{equation}
M_{(8|1)\times(8|1)}=\left(\begin{array}{c:c|c}
\so(2,4) & 0 & \overline{Q}\\ 
\hdashline
0 & \su(2)\oplus\su(2)& 0\\ 
\hline
Q& 0 & \u(1)_R
\end{array}\right),
\end{equation}
where the dashed lines split the algebras corresponding to the subspaces $\AdS{5}$ and $\T{1,1}$, while the solid lines split the $M_{8\times 8}$ and $M_{1\times 1}$ bosonic blocks.

The basis of $\so(2,4)\oplus\su(2)\oplus\su(2)\oplus\u(1)$ that we will consider is composed of $\so(2,4)$ generators denoted by $\G_\mu$, $\G_5$, $\SOM{\mu\nu}$ and $\SOM{\mu5}$, $\mu=0,1,2,3$, which, when written as supermatrices become 
\begin{equation}
\G_\mu=\left(\begin{array}{c:cc|c}
\g_\mu &  &  & \\ 
\hdashline
 & 0 &  & \\ 
 &  & 0 & \\ 
\hline
 &  &  &0
\end{array}\right)\!\!,\quad
\G_5=\left(\begin{array}{c:cc|c}
\g_5 &  &  & \\ 
\hdashline
 & 0 &  & \\ 
 &  & 0 & \\ 
\hline
 &  &  &0
\end{array}\right)\!\!,\quad
\SOM{\mu\nu}=\left(\begin{array}{c:cc|c}
m_{\mu\nu} &  &  & \\ 
\hdashline
 & 0 &  & \\ 
 &  & 0 & \\ 
\hline
 &  &  &0
\end{array}\right)\!\!,\quad
\SOM{\mu5}=\left(\begin{array}{c:cc|c}
m_{\mu5} &  &  & \\ 
\hdashline
 & 0 &  & \\ 
 &  & 0 & \\ 
\hline
 &  &  &0
\end{array}\right)\!\!,
\end{equation}
and $\su(2)\oplus\su(2)\oplus\u(1)$ generators denoted by $\EX{a},\EY{a}$, $a=1,2,3$ and $\EM$, with   supermatrices 
\begin{equation}
\EX{a}=-\frac{i}{2}\left(\begin{array}{c:cc|c}
0 &  &  & \\ 
\hdashline
 & \es{a} &  & \\ 
 &  & 0 & \\ 
\hline
 &  &  &0
\end{array}\right)\!\!,\quad
\EY{a}=-\frac{i}{2}\left(\begin{array}{c:cc|c}
0 &  &  & \\ 
\hdashline
 & 0 &  & \\ 
 &  & \es{a}& \\ 
\hline
 &  &  &0
\end{array}\right)\!\!,\quad
\EM=-\frac{i}{2}\left(\begin{array}{c:cc|c}
0 &  &  & \\ 
\hdashline
 & 0 &  & \\ 
 &  & 0 & \\ 
\hline
 &  &  & 1
\end{array}\right)\!\!,\quad
\end{equation}
where $\g_\mu$, $\g_5$, $m_{\mu\nu}$ and $m_{\mu5}$ are the fifteen $4\times 4$ matrices for the generators of isometries of $\AdS{5}$ (detailed in \autoref{ApSO24}) and $\es{a}$ are the conventional $2\times 2$ Pauli matrices of $\su(2)$. The commutation rules and supertraces are then
\begin{equation}
\left[\SOM{ij},\SOM{k\ell}\right]=\Et{i\ell}\SOM{jk}+\Et{jk}\SOM{i\ell}-\Et{ik}\SOM{j\ell}-\Et{j\ell}\SOM{ik},
\end{equation}
where $i,j,k,\ell=0,1,\dots,5$, and 
\begin{equation}
\begin{gathered}
\left[\EX{a},\EX{b}\right]=\eabc{a}{b}{c}\EX{c},\qquad\left[\EY{a},\EY{b}\right]=\eabc{a}{b}{c}\EY{c},\\
\Str\left(\EX{a}\EX{b}\right)=-\frac{1}{2}\d_{ab},\qquad\Str\left(\EY{a}\EY{b}\right)=-\frac{1}{2}\d_{ab},\\
\Str\left(\EM\EM\right)=\frac{1}{4},
\end{gathered}
\end{equation}
with $\Et{ij}=\diag\left(-,+,+,+,-,+,+,+,+,+\right)$. 

The algebra for the global symmetry of the $\AdS{5}$ space is
\begin{equation}
\so(2,4)=\so(1,4)\oplus\frac{\so(2,4)}{\so(1,4)}, 
\end{equation}
with basis 
\begin{equation}
\frac{\so(2,4)}{\so(1,4)}=\text{span}\left\{\EK{m}\right\},\qquad m=0,1,2,3,4,
\end{equation}
where 
\begin{equation}
\label{EKAdS5}
\EK{0}=\frac{1}{2}\G_0,\quad\EK{1}=\frac{1}{2}\G_1,\quad\EK{2}=\frac{1}{2}\G_2,\quad\EK{3}=\frac{1}{2}\G_3,\quad\EK{4}=\frac{1}{2}\G_5,
\end{equation}
and
\begin{equation}
\Str\left(\EK{m}\EK{n}\right)=\eta_{mn},\qquad m,n=0,1,2,3,4.
\end{equation}
The $\so(1,4)$ generators are $\left\{\SOM{01},\SOM{02},\SOM{03},\SOM{12},\SOM{13},\SOM{23},\SOM{05},\SOM{15},\SOM{25},\SOM{35}\right\}$, and an appropriate coset representative for $\AdS{5}$ is
\begin{equation}\label{gAdS5}
g_{\AdS{5}}=\exp\left(\x{0}\Ep{0}+\x{1}\Ep{1}+\x{2}\Ep{2}+\x{3}\Ep{3}\right)\exp\left(\log z\ED\right),
\end{equation}
where 
\begin{equation}
\ED=\frac{1}{2}\G_5,\qquad\Ep{\mu}=\frac{1}{2}\G_\mu+\SOM{\mu5}\,,\qquad \mu=0,1,2,3.
\end{equation}

The $\T{1,1}$ space can be written as the coset in
\begin{equation}
\su(2)\oplus\su(2)\oplus\u(1)=\u(1)\oplus\u(1)\oplus\frac{\su(2)\oplus\su(2)\oplus\u(1)}{\u(1)\oplus\u(1)}.
\end{equation}
with basis 
\begin{equation}
\frac{\su(2)\oplus\su(2)\oplus\u(1)}{\u(1)\oplus\u(1)}=\text{span}\left\{\EK{m}\right\},\qquad m=5,\dots,9,
\end{equation}
where 
\begin{equation}
\label{EKT11}
\begin{gathered}
\EK{5}=\sqrt{\frac{2}{3}}\EX{1},\quad\EK{6}=\sqrt{\frac{2}{3}}\EX{2},\quad\EK{7}=\sqrt{\frac{2}{3}}\EY{1},\\
\EK{8}=\sqrt{\frac{2}{3}}\EY{2},\quad\EK{9}=\frac{2}{3}\EH,
\end{gathered}
\end{equation}
with 
\begin{equation}
\EH=\EX{3}-\EY{3}+\EM.
\end{equation}
We also have
\begin{equation}
\Str\left(\EK{m}\EK{n}\right)=-\frac{1}{3}\d_{mn},\qquad m,n=5,\dots,9.
\end{equation}
The generators of $\u(1)\oplus\u(1)$ are $\left\{\ET{1},\ET{2}\right\}$ with
\begin{equation}
\ET{1}=\EX{3}+\EY{3},\qquad\ET{2}=\EX{3}-\EY{3}+4\EM,
\end{equation}
where $T_1$ generates the original $U(1)$ in \eqref{isometry}. An appropriate coset representative is then 
\begin{equation}\label{gT11}
g_{\T{1,1}}=g=\exp\left(\ph{1}\EX{1}+\ph{2}\EX{2}+2\ph{3}\EM\right)\exp\left(\th{1}\EX{2}+(\th{2}+\pi)\EY{2}\right).
\end{equation}
The coset representative that will allow use for $\AdS{4}\times\T{1,1}$ is then 
\begin{equation}
g=g_{\AdS{5}}\times g_{\T{1,1}}.
\end{equation}

The projector $P_2$ on $\algi{2}$ can be defined as
\begin{equation}
\label{P2T11}
\Pb{X}=\sum_{m=0}^4\frac{\Str\left(\EK{m}X\right)}{\Str\left(\EK{m}\EK{m}\right)}\EK{m}-\frac{1}{3}\sum_{m=5}^9\frac{\Str\left(\EK{m}X\right)}{\Str\left(\EK{m}\EK{m}\right)}\EK{m}.
\end{equation}
Applied to $A=g^{-1}dg$, the Maurer-Cartan one-form, we get 
\begin{equation}
\Pb{A}=\Em{m}\EK{m},\qquad m=0,1,\dots,9,
\end{equation}
with
\begin{equation}
\label{EmAdS5T11}
\begin{gathered}
\Em{0}=\frac{\dx{0}}{z},\quad\Em{1}=\frac{\dx{1}}{z},\quad\Em{2}=\frac{\dx{2}}{z},\quad\Em{3}=\frac{\dx{3}}{z},\quad\Em{4}=\frac{dz}{z},\\
\Em{5}=\frac{1}{\sqrt{6}}\sin\th{1}\dph{1},\quad\Em{6}=-\frac{1}{\sqrt{6}}\dth{1},\quad\Em{7}=-\frac{1}{\sqrt{6}}\sin\th{2}\dph{2},\quad\Em{8}=-\frac{1}{\sqrt{6}}\dth{2},\\
\Em{9}=-\frac{1}{3}\left(\cos\th{1}\dph{1}+\cos\th{2}\dph{2}+\dph{3}\right).
\end{gathered}
\end{equation}
Then, we can compute the $\AdS{5}\times\T{1,1}$ metric from
\begin{equation}
\label{APAT11}
\Str\left(A\,\Pb{A}\right)=\Em{m}\Str\left(A\EK{m}\right),\quad m=0,\dots,9\,,
\end{equation}
to get
\begin{equation}
ds^2=ds^2_{\AdS{5}}+ds^2_{\T{1,1}},
\end{equation}
where
\begin{equation}
ds^2_{\AdS{5}}=\frac{1}{z^2}\left(-\dxs{0}+\dxs{1}+\dxs{2}+\dxs{3}+dz^2\right), 
\end{equation}
and 
\begin{equation}
\label{T11metric}
\begin{gathered}
ds^2_{\T{1,1}}=\frac{1}{6}\left(\dth{1}^2+\sin^2\!\th{1}\dph{1}^2\right)+\frac{1}{6}\left(\dth{2}^2+\sin^2\!\th{2}\dph{2}^2\right)\\
+\frac{1}{9}\left(\cos\th{1}\dph{1}+\cos\th{2}\dph{2}+\dph{3}\right)^2,
\end{gathered}
\end{equation}
where $\left(\th{1},\ph{1}\right)$ and $\left(\th{2},\ph{2}\right)$ parametrize the two spheres of $\T{1,1}$ and $0\leq\ph{3}\leq2\pi$.

The metric \eqref{T11metric} was first obtained in \cite{Candelas1990} and describes the basis of a six-dimensional cone. It can be understood as the intersection of a cone and a sphere in $\CC^4$ such that its topology is $\Sph{2}\times\Sph{3}$, and that the metric is a $U(1)$ bundle over $\Sph{2}\times\Sph{2}$.  Besides that, $SO(4)\cong SU(2)\times SU(2)$ acts transitively on $\Sph{2}\times\Sph{3}$ and $U(1)$ leaves each point of it fixed so that $\T{1,1}$ is described by the coset $\left(SU(2)\times SU(2)\right)/U(1)$. 

\section{Yang-Baxter deformed backgrounds \label{news}}
In this section we present some $r$-matrices satisfying the CYBE and build the corresponding 
deformed  background identifying its gravity dual. As mentioned before, the background can be deformed partially by choosing generators on each subspace.
The bosonic Yang-Baxter deformed action is \cite{Delduc2014}
\begin{equation}
\label{YBdef}
S=-\frac{1}{2}\int\dds\left(\gab{\a\b}-\eab{\a\b}\right)\Str\left(A_{\a}\Pb{J_{\b}}\right),
\end{equation}
where $A=g^{-1}dg\in\algg$ , $\g^{\a\b}$ is the worldsheet metric and $\eab{\a\b}$ is the Levi-Civita symbol. $P_2$ was defined in \eqref{P2T11} and the deformed current one-form is
\begin{equation}\label{J1}
J=\frac{1}{1-2\eta R_g\o P_2}A,
\end{equation}
where $\eta$ is the deformation parameter. The dressed $R$ operator $R_g$ is defined as
\begin{equation}
\label{Rgdef}
\Rg{M}=\Adgi\o R\o\Adg\left(M\right)=g^{-1}R(gMg^{-1})g.
\end{equation}
Moreover, we can compute $\Pb{J}$ in \eqref{YBdef} by defining the action of $P_2$ as 
\begin{equation}\label{PJK}
\Pb{A}=\Em{m}\EK{m}, \qquad \Pb{J}=\jm{m}\EK{m}.
\end{equation}
The coefficients $\jm{m}$ can be calculated from
\begin{equation}\label{Emn}
\jm{m}=\Em{n}\Cmn{n}{m},
\end{equation}
where the matrix components $\Cmn{m}{n}$ are those of
\begin{equation}\label{Cmatrix}
\mathbf{C}=\left(\mathbf{I}-2\eta\mathbf{\Lambda}\right)^{-1}.
\end{equation}
The matrix $\mathbf{\Lambda}$ has components defined as 
\begin{equation}\label{Lmatrix}
\Pb{\Rg{\EK{m}}}=\Lmn{m}{n}\EK{n}.
\end{equation}
Then, from \eqref{YBdef}, we can read off the metric and the B-field as \cite{Rado2020}
\begin{equation}\label{Gmn}
ds^2=\Str\left(A\,\Pb{J}\right)=\jm{m}\Str\left(A\EK{m}\right)=\Em{m}\Cmn{m}{n}\Str\left(A\EK{n}\right),
\end{equation} 
\begin{equation}\label{Bmn}
B=\Str\left(A\wedge\Pb{J}\right)=-\jm{m}\wedge\Str\left(A\EK{m}\right)=\Em{m}\Cmn{m}{n}\wedge\Str\left(A\EK{n}\right).
\end{equation}

The three-parameter $\b$-deformed of $\T{1,1}$ was obtained in \cite{Crichigno2014} by a Yang-Baxter deformation and in \cite{Catal2006} by a TsT transformation in perfect agreement. In this case the $r$-matrix was
\begin{equation}
\label{betadef}
r=\mu_1 \EX{3}\wedge\EM+\mu_2\EM\wedge\EY{3}+\mu_3\EX{3}\wedge\EY{3}.
\end{equation}
In the following subsections we will introduce two more $r$-matrices and the corresponding deformations they produce.
\subsection{Dipole Deformed Klebanov-Witten Theory \label{dipole}}
Let us first consider an Abelian $r$-matrix like
\begin{equation}
\label{rdipole}
r=\Ep{2}\wedge\left(\mu_1\EX{3}+\mu_2\EY{3}+\mu_3\EM\right),
\end{equation}
where $\EX{3}$, $\EY{3}$ and $\EM$ are the Cartan generators of $\su(2)\oplus\su(2)\oplus\u(1)$ and $\mu_i$, $i=1,2,3$, are the deformation parameters \footnote{The deformation parameter $\eta$ can always be absorbed in the $r$-matrix such that it is present in the $\mu_i$'s.}. In this case \eqref{rdipole} combines generators of both subspaces, which will lead to a deformation of the entire $\AdS{5}\times\T{1,1}$ background. 
The nonzero components of $\Lmn{m}{n}$ in \eqref{Lmatrix} are 
\begin{equation}
\begin{gathered}
\Lmn{3}{5}=-\Lmn{5}{3}=-\frac{1}{\sqrt{6}}\frac{\mu_1\sin\th{1}}{z},\\
\Lmn{3}{7}=-\Lmn{7}{3}=\frac{1}{\sqrt{6}}\frac{\mu_2\sin\th{2}}{z},\\
\Lmn{3}{9}=-\Lmn{9}{3}=\frac{1}{6}\frac{\left(\mu_3-2\mu_1\cos\th{1}-2\mu_2\cos\th{2}\right)}{z},
\end{gathered}
\end{equation}
while the nonzero elements of $\Cmn{m}{n}$, from \eqref{Cmatrix}, are
\begin{equation}
\begin{gathered}
\Cmn{0}{0}=\Cmn{1}{1}=\Cmn{2}{2}=\Cmn{4}{4}=\Cmn{6}{6}=\Cmn{8}{8}=1,\\
\Cmn{3}{3}=\M,\\
\Cmn{5}{3}=-\Cmn{3}{5}=\M f_1,\\
\Cmn{7}{3}=-\Cmn{3}{7}=-\M f_2,\\
\Cmn{9}{3}=-\Cmn{3}{9}=\M f_3,\\
\Cmn{5}{5}=\M\left(1+f_2^2+f_3^2\right),\\
\Cmn{7}{5}=\Cmn{5}{7}=\M f_1 f_2,\\
\Cmn{9}{5}=\Cmn{5}{9}=-\M f_1 f_3,\\
\Cmn{7}{7}=\M\left(1+f_1^2+f_3^2\right),\\
\Cmn{9}{7}=\Cmn{7}{9}=\M f_2 f_3,\\
\Cmn{9}{9}=\M\left(1+f_1^2+f_2^2\right),
\end{gathered}
\end{equation}
where
\begin{equation}
\begin{gathered}
\M^{-1}=1+f_1^2+f_2^2+f_3^2,
\end{gathered}
\end{equation}
with
\begin{equation}
\begin{gathered}
f_1=\sqrt{\frac{2}{3}}\frac{\mu_1\sin\th{1}}{z},\quad f_2=\sqrt{\frac{2}{3}}\frac{\mu_2\sin\th{2}}{z},\\
f_3=\frac{\mu_3-2\mu_1\cos\th{1}-2\mu_2\cos\th{2}}{3z}.
\end{gathered}
\end{equation}
The deformed metric can be obtained from \eqref{Gmn}
\begin{equation}
\label{metricdipole3}
\begin{gathered}
ds^2=\frac{1}{z^2}\left(-\dxs{0}+\dxs{1}+\dxs{2}+\M\dxs{3}+dz^2\right)\\
+\frac{1}{6}\left(\dth{1}^2+\M\left(1+f_2^2+f_3^2\right)\sin^2\!\th{1}\dph{1}^2\right)+\frac{1}{6}\left(\dth{2}^2+\M\left(1+f_1^2+f_3^2\right)\sin^2\!\th{2}\dph{2}^2\right)\\
+\frac{\M}{9}\left(1+f_1^2+f_2^2\right)\left(\cos\th{1}\dph{1}+\cos\th{2}\dph{2}+\dph{3}\right)^2\\
+\frac{\sqrt{6}\M}{9}f_3\left(f_1\sin\th{1}\dph{1}+f_2\sin\th{2}\dph{2}\right)\left(\cos\th{1}\dph{1}+\cos\th{2}\dph{2}+\dph{3}\right)\\
-\frac{\M}{3}f_1f_2\sin\th{1}\sin\th{2}\dph{1}\dph{2},
\end{gathered}
\end{equation}
and the $B$-field from \eqref{Bmn},
\begin{equation}
\label{Bdipole}
\begin{gathered}
B=-\frac{\M}{3z}\left(2f_3\cos\th{1}-\sqrt{6}f_1\sin\th{1}\right)\dx{3}\wedge\dph{1}\\
-\frac{\M}{3z}\left(2f_3\cos\th{2}-\sqrt{6}f_2\sin\th{2}\right)\dx{3}\wedge\dph{2}\\
-\frac{2\M}{3z}f_3\dx{3}\wedge\dph{3}.
\end{gathered}
\end{equation}
It is worth mentioning that the choice of generators in \eqref{rdipole} is dictated by the place where we want put the two-tori from the TsT perspective. In the present case we have one coordinate in $\AdS{5}$ and a combination of the $U(1)$'s in $\T{1,1}$. The resulting metric \eqref{metricdipole3} has  deformations along the $\x{3}$-direction in $\AdS{5}$ and along the angles $\ph{1},\ph{2}$ and $\ph{3}$  in $\T{1,1}$.
\subsection{Nonrelativistic Klebanov-Witten Theory \label{Sch5}}
In order to construct this deformation we must write the $\AdS{5}$ space in light-cone coordinates. Thus, the coset representative is now 
\begin{equation}\label{gAdS5LC}
g_{\AdS{5}}=\exp\left(\xm\Epm+\xp\Epp+\x{1}\Ep{1}+\x{2}\Ep{2}\right)\exp\left(\log z\ED\right),
\end{equation}
with
\begin{equation}
\Ep{\pm}=\frac{1}{\sqrt{2}}\left(\Ep{0}\pm\Ep{3}\right),\qquad x_{\pm}=\frac{1}{\sqrt{2}}\left(\x{0}\pm\x{3}\right), 
\end{equation}
while for the $\T{1,1}$ we keep the same form as in \eqref{gT11}.
The $\AdS{5}$ metric is then
\begin{equation}
ds^2=\frac{1}{z^2}\left(-2\dxp\dxm+\dxs{1}+\dxs{2}+dz^2\right), 
\end{equation}
while the $\T{1,1}$ metric is given by \eqref{T11metric}.

Let us now consider the $r$-matrix \eqref{rdipole} with $\Ep{2}$ replaced by $\Epm$ \footnote{In this case we identify $\xm\sim\xm+2\pi r^{-}$, such that $p^{-}=i\partial_{\xm}$ can be interpreted as the number operator $p^{-}=N/r^{-}$. Moreover, if we consider $\xp$ to be the time then $p^{+}$ is the energy \cite{Hartnoll2008}.}
\begin{equation}
\label{rSch5}
r=\Epm\wedge\left(\mu_1\EX{3}+\mu_2\EY{3}+\mu_3\EM\right),
\end{equation}
where $\EX{3}$, $\EY{3}$ and $\EM$ are Cartan generators of the algebra. Taking the same steps as in the previous case we find that the nonzero components of $\Lmn{m}{n}$ are 
\begin{equation}
\begin{gathered}
\Lmn{0}{5}=\Lmn{5}{0}=\frac{1}{2\sqrt{3}}\frac{\mu_1\sin\th{1}}{z},\\
\Lmn{0}{7}=\Lmn{7}{0}=-\frac{1}{2\sqrt{3}}\frac{\mu_2\sin\th{2}}{z},\\
\Lmn{0}{9}=\Lmn{9}{0}=\frac{1}{6\sqrt{2}}\frac{\mu_3-2\mu_1\cos\th{1}-2\mu_2\cos\th{2}}{z},\\
\Lmn{3}{5}=-\Lmn{5}{3}=\frac{1}{2\sqrt{3}}\frac{\mu_1\sin\th{1}}{z},\\
\Lmn{3}{7}=-\Lmn{7}{3}=-\frac{1}{2\sqrt{3}}\frac{\mu_2\sin\th{2}}{z},\\
\Lmn{3}{9}=-\Lmn{9}{3}=\frac{1}{6\sqrt{2}}\frac{\mu_3-2\mu_1\cos\th{1}-2\mu_2\cos\th{2}}{z},
\end{gathered}
\end{equation}
while the nonzero elements of $\Cmn{m}{n}$ are now
\begin{equation}
\begin{gathered}
\Cmn{1}{1}=\Cmn{2}{2}=\Cmn{4}{4}=\Cmn{5}{5}=\Cmn{6}{6}=\Cmn{7}{7}=\Cmn{8}{8}=\Cmn{9}{9}=1,\\
\Cmn{0}{0}=1+f_1^2+f_2^2+f_3^2,\\
\Cmn{3}{0}=-\Cmn{0}{3}=-\left(f_1^2+f_2^2+f_3^2\right),\\
\Cmn{5}{0}=\Cmn{0}{5}=f_1,\\
\Cmn{7}{0}=\Cmn{0}{7}=-f_2,\\
\Cmn{9}{0}=\Cmn{0}{9}=f_3,\\
\Cmn{3}{3}=1-\left(f_1^2+f_2^2+f_3^2\right),\\
\Cmn{5}{3}=-\Cmn{3}{5}=-f_1,\\
\Cmn{7}{3}=-\Cmn{3}{7}=f_2,\\
\Cmn{9}{3}=-\Cmn{3}{9}=-f_3,
\end{gathered}
\end{equation}
where
\begin{equation}
\begin{gathered}
f_1=\frac{1}{\sqrt{3}}\frac{\mu_1\sin\th{1}}{z},\quad f_2=\frac{1}{\sqrt{3}}\frac{\mu_2\sin\th{2}}{z},\\
f_3=\frac{1}{3\sqrt{2}}\frac{\mu_3-2\mu_1\cos\th{1}-2\mu_2\cos\th{2}}{z}.
\end{gathered}
\end{equation}
The deformed metric is then
\begin{equation}
\label{metricSch5}
ds^2=\frac{1}{z^2}\left(-2\dxp\dxm+\dxs{1}+\dxs{2}+dz^2\right)-2\M\frac{\dxps}{z^2}+ds^2_{\T{1,1}},
\end{equation}
where now 
\begin{equation}
\M=f_1^2+f_2^2+f_3^2, 
\end{equation}
while the deformed $B$-field is 
\begin{equation}
\label{BSch5}
\begin{gathered}
B=-\frac{2}{3z}\left(\sqrt{2}f_3\cos\th{1}+\sqrt{3}f_1\sin\th{1}\right)\dxp\wedge\dph{1}\\
-\frac{2}{3z}\left(\sqrt{2}f_3\cos\th{2}+\sqrt{3}f_2\sin\th{2}\right)\dxp\wedge\dph{2}\\
-\frac{2\sqrt{2}}{3z}f_3\dxp\wedge\dph{3}.
\end{gathered}
\end{equation}
The first two terms in \eqref{metricSch5} is the metric of a Schr\"odinger spacetime\footnote{{The Schr\"odinger symmetry is the maximal symmetry group of the free Schr\"odinger equation. It is the nonrelativistic version of the conformal algebra \cite{Hagen1972,Niederer1972}. This symmetry is realized geometrically as Schr\"odinger spacetimes.}}. 
The choice of generators in \eqref{rSch5} is very similar to the one in \eqref{rdipole}. Now, however, the two-tori defined by the TsT transformation takes the $\xm$ coordinate and a combination of the internal $U(1)$'s in ${\T{1,1}}$ and does not introduce any noncommutativity in the dual field theory. {The metric \eqref{metricSch5} coincide with the $\Sch{5}\times\T{1,1}$ obtained in \cite{Bobev2009a} for $\mu_1=n_1/2,\,\mu_2=n_2/2$ and $\mu_3=-n_3$, where $n_i$ $(i=1,2,3)$ are the deformation parameters.

The Schr\"odinger spacetime in \eqref{metricSch5} has dynamical exponent two \cite{Son2008,Balasubramanian2008} \footnote{{The dynamical $z$ factor is the exponent in the power of the radial direction in the $z^{-2z}\dxps$ term. To have Schr\"odinger symmetry we must have $z=2$. The relativistic symmetry corresponds to $z=1$.}}. Schr\"odinger backgrounds with dynamical exponent $z$ are argued to be integrable for $z=1,2,3$, and non-integrable for $z=4,5,6$ \cite{Giataganas2014}. It has been argued that there are several nonrelativistic gravity duals with Schr\"odinger symmetry \cite{Hartnoll2008}. The number of $\Sch{5}\times\T{1,1}$ spaces is equal to the degeneracy of a scalar harmonic function $\Phi_{(0)}^{(\ell_1,\ell_2,r)}$ on $\T{1,1}$, 
with quantum numbers $(\ell_1,\ell_2,r)$, for which the Laplace-Beltrami equation is $-\Lap\Phi_{(0)}^{(\ell_1,\ell_2,r)}=\l_{(\ell_1,\ell_2,r)}\Phi_{(0)}^{(\ell_1,\ell_2,r)}$ with $\l_{(\ell_1,\ell_2,r)}=6\left(\ell_1\left(\ell_1+1\right)+\ell_2\left(\ell_2+1\right)-r^2/8\right)$\footnote{{ The harmonic function is denoted in general as $\Phi_{(q)}^{(\ell_1,\ell_2,r)}$, where $\ell_1,\ell_2$ are labels for the $SU(2)$'s, and $r$ and $q$ are $U(1)$ charges \cite{Gubser1999,Ceresole2000}.}}. Since $\ell_1,\ell_2$ label standard spherical harmonics on the $\Sph{2}$'s of $\T{1,1}$ the multiplicities of $\Phi_{(0)}^{(\ell_1,\ell_2,r)}$ are $(2|\ell_1|+1)$ and $(2|\ell_2|+1)$ \cite{Hartnoll2008}. Our background has\footnote{{In \cite{Bobev2009a}, the harmonic function $\Phi$ is defined as the non-negative length square of the Killing vector $\K$ on $\T{1,1}$, $\Phi=\norm{\K}^2=g_{ij}\K^i\K^j$ with $i,j=1,2,3$, where $\K=\left(\mu_1\partial_{\ph{1}},\mu_2\partial_{\ph{2}},\mu_3\partial_{\ph{3}}\right)$.}} $\Phi=2\M z^2$  and $-\Lap\Phi_{(0)}^{(\ell_1,\ell_2,r)}=12\Phi_{(0)}^{(\ell_1,\ell_2,r)}$, so that $\left(\ell_1,\ell_2,r\right)$ takes two values, $\left(1,0,0\right)$ and $\left(0,1,0\right)$. Then the total degeneracy of $\Phi_{(0)}$ is six so that we have a family of six $\Sch{5}\times\T{1,1}$ spacetimes \cite{Hartnoll2008}. This kind of spacetimes was recently studied in \cite{Golubtsova2020a,Golubtsova2020}.
\section{Conclusions \label{conclusions}}
In this paper we have derived the metric and the $B$-field for the gravity duals of the dipole-deformed and the nonrelativistic Klebanov-Witten theory as Yang-Baxter deformations. We made use of an extended coset description of $\AdS{5}\times\T{1,1}$ which simplified the computation of the undeformed background and its deformation. We considered two abelian $r$-matrices with three-parameter satisfying the classical Yang-Baxter equation. The first $r$-matrix was composed by a momentum generator in $AdS$ and a combination of the three $U(1)$'s generators of the internal space which lead to the gravity dual of the dipole-deformed Klebanov-Witten theory which should be obtained by TsT transformation of the $\AdS{5}\times\T{1,1}$ background. In second case we have also a momentum operator in $AdS$ and a combination of the three $U(1)$ generators in $\T{1,1}$. It produced the $\Sch{5}\times\T{1,1}$ background which, having Schr\"odinger symmetry, corresponds to the nonrelativistic Klebanov-Witten theory \cite{Balasubramanian2008}. 

The next step is to compute the RR fields of the deformed backgrounds. To get them we have to consider the fermionic sector as in \cite{Kyono2016}. The fact that we have not included the fermionic sector of the supercoset does not mean that we are unable to check the supergravity equations for the new backgrounds. Since the $r$-matrices that we used in the bosonic background are abelian they satisfy trivially the unimodularity condition, which is a  sufficient for the background to satisfy the supergravity equations \cite{Wulff2016,Arutyunov2016,Borsato2016}.


Another interesting case which deserves further study is the dual of the dipole deformation of $\N=1$ $SU(N)\times SU(N)$ Yang-Mills theory as well as its nonrelativistic limits.

\appendix
\section{A basis for \texorpdfstring{$\so(2,4)$}{so(2,4)} algebra \label{ApSO24}} 

Let us choose the following representation for $\gamma_\mu$
\begin{equation}
\begin{gathered}
\g_0=\begin{pmatrix}
 0 &  \es{3}  \\ 
 -\es{3} & 0 \\  
\end{pmatrix},\quad
\g_1=\begin{pmatrix}
 0 &  -i\es{2}  \\ 
 i\es{2}   & 0 \\  
\end{pmatrix},\\
\g_2=\begin{pmatrix}
 0 &  i\es{1}  \\ 
 -i\es{1}  & 0 \\  
\end{pmatrix},\quad
\g_3=\begin{pmatrix}
 0 &  I_2  \\ 
 -I_2 & 0 \\  
\end{pmatrix},
\end{gathered}
\end{equation}
and 
\begin{equation}
\g_5=\begin{pmatrix}
I_2 & 0  \\ 
0 & -I_2 \\  
\end{pmatrix}.
\end{equation}
We can also define
\begin{equation}
m_{\mu\nu}=\frac{1}{4}\left[\g_\mu,\g_\nu\right],\quad m_{\mu 5}=\frac{1}{4}\left[\g_\mu,\g_5\right], 
\end{equation}
and
\begin{align}\nonumber
p_\mu&=\frac{1}{2}\g_\mu-m_{\mu 5},\\ \nonumber
k_\nu&=\frac{1}{2}\g_\mu+m_{\mu 5},\\
D&=\frac{1}{2}\g_5.
\end{align}
The conformal algebra $SO(2,4)$ is then
\begin{align}\nonumber
\left[m_{\mu\nu},m_{\rho\sigma}\right]&=\eta_{\mu\sigma}m_{\nu\rho}+\eta_{\nu\rho}m_{\mu\sigma}-\eta_{\mu\rho}m_{\nu\sigma}-\eta_{\nu\sigma}m_{\mu\rho},\\
\nonumber
\left[m_{\mu\nu},D\right]&=0,\\
\nonumber
\left[D,p_\mu\right]&=p_\mu,\\
\label{conformal4}
\left[D,k_\mu\right]&=-k_\mu,\\
\nonumber
\left[k_\mu,p_\nu\right]&=2\eta_{\mu\nu}D+2m_{\mu\nu},\\
\nonumber
\left[m_{\mu\nu},p_\rho\right]&=-\eta_{\mu\rho}P_\nu+\eta_{\nu\rho}P_\mu,\\
\nonumber
\left[m_{\mu\nu},k_\rho\right]&=-\eta_{\mu\rho}k_\nu+\eta_{\nu\rho}k_\mu.
\end{align}
\acknowledgments
The work of L. Rado was supported by CAPES. The work of V.O. Rivelles was supported by FAPESP grant 2019/21281-4.

\bibliographystyle{JHEP}
\bibliography{Biblioteca}
\end{document}